\documentclass[a4paper]{jpconf}
\usepackage[cp1252]{inputenc}
\usepackage{iopams}
\begin{document}
\title{Monte-Carlo Geant4 numerical simulation of experiments at 247--MeV proton microscope}

\author{A~V~Kantsyrev$^{1}$, A~V~Skoblyakov$^{1}$, A~V~Bogdanov$^{1}$, A~A~Golubev$^{1}$, N~S~Shilkin$^{2}$, D~S~Yuriev$^{2}$ and V~B~Mintsev$^{2}$}

\address{$^1$ State Scientific Center of the Russian Federation ``Institute for Theoretical and Experimental Physics'',  National Research Center ``Kurchatov Institute'', Bolshaya Cheremushkinskaya 25, Moscow 117218, Russia}
\address{$^2$ Institute of Problems of Chemical Physics of the Russian Academy of Sciences, Academician Semenov Avenue 1, Chernogolovka, Moscow Region 142432, Russia}

\ead{kantsyrev@itep.ru}

\begin{abstract}
A radiographic setup for an investigation of fast dynamic processes with areal density of targets up to 5~g/cm$^2$ is under development on the basis of high-current proton linear accelerator at the Institute for Nuclear Research (Troitsk, Russia). A virtual model of the proton microscope developed in a software toolkit Geant4 is presented in the article. Full-scale Monte-Carlo numerical simulation of static radiographic experiments at energy of a proton beam 247~MeV was performed. The results of simulation of proton radiography experiments with static model of shock-compressed xenon are presented. The results of visualization of copper and polymethyl methacrylate step wedges static targets also described.  
\end{abstract}
\bibliographystyle{iopart-num}
\section{Introduction}

Interest in physical properties of states of matter with high energy density is defined by numerous cross-disciplinary scientific tasks and a number of important practical applications \cite{Fortov09}. Theoretical models of thermodynamic and transport properties of matter were developed for states with weak interparticle interaction, i.e. states at high temperature or at high density. In the region of strong interparticle interaction the matter becomes non-ideal and applicability of strict theoretical approaches is limited to lack of small parameters. For this reason experimental information becomes extremely important for verification of the used models and validation of assumptions. States of matter with strong interparticle interaction is generated by different experimental methods, including diamond anvils, explosive driven and light-gas guns accelerators, high-power lasers and charged particles beams, explosive generators, magnetic cumulative generators, pulse power devices. Maximum densities of energy, pressures and temperatures are generated by dynamic methods, the studied states are created on a short period. Methods of determination of kinematic parameters of gas-dynamic processes \cite{Gor55Kin, Alt65Kin,FritzKin}, thermodynamic \cite{BurrowsTerm, graham65term, model65Term} and transport \cite{IvanovTr, ShilkinTr, ZaporTr} properties of states generated in a dynamic experiment were developed. Density in a dynamic experiment is usually determined from conservation laws by the measured values of kinematic parameters of motion of shock waves and contact discontinuities \cite{Zeld} or method of x-ray radiography \cite{rentrad}. Accuracy of density determination of shock compressed inert gases by pulse x-ray radiography strongly depends on density of gas, a design of the generator and conditions of carrying out an experiment. The density of shock compressed xenon in a diaphragm shock tube was measured with accuracy 4--6\% in work \cite{Bushman1975}, the error of density measurement of the shock-compressed argon was estimated as 8\% in \cite{Bespalov1975}. Iisentropic compression of helium in high explosive cylindrical two-cascade generator was reported in \cite{Zhernokletov}, density was measured with accuracy 5\%.The results on compression of helium and deuterium in hemispherical generators were published in work \cite{Mochalov2012}, the density was determined with accuracy up to 15--20\% at final stages of compression. High-energy proton radiography is another method of density measurement \cite{West}. Application of magnetic optics considerably improved spatial resolution and contrast sensitivity of proton radiography \cite{King99}, spatial resolution of 60--120 $\mu$m in targets with areal density up to 20~g/cm$^2$ and proton energy 800~MeV \cite{Kantsyrev2014} was achieved at PUMA proton microscope at TWAC--ITEP facility. Results of a numerical simulation of an experiment on determination of density of the shock-compressed xenon on the proton microscope at the Institute for Nuclear Research of Russian Academy of Sciences (Troitsk, Russia) with energy of 247~MeV are presented below.

\section{The model of proton radiography facility}

The model of proton microscope was developed in the software toolkit Geant4 \cite{Geant} (figure~1), which consists of a vacuum beamline, the magnetic quadrupole lenses, the detector and a target with areal density up to 5~g/cm$^2$. The scheme of the facility model is similar to earlier developed PUMA microscope at ITEP \cite{Kantsyrev2014}. Aluminum was chosen as material of the beamline for reduction of activation, the thickness of the beamline walls is 2 mm. The magnetic quadrupole lenses are defined in the form of rectangular steel parallelepipeds with the internal diameter corresponding to an aperture of lenses. The spatial distribution of a magnetic field in lenses was set as ideal quadrupole. The detector for registering of protons was defined by a set of sensitive elements in vacuum with total number of $2000 \times 2000$ elements. Multiple Coulomb scattering, ionization energy losses and nuclear interactions are included in physics of processes that took place in the target.

\begin{figure}[t]
\begin{center}
\includegraphics[width=0.8\columnwidth]{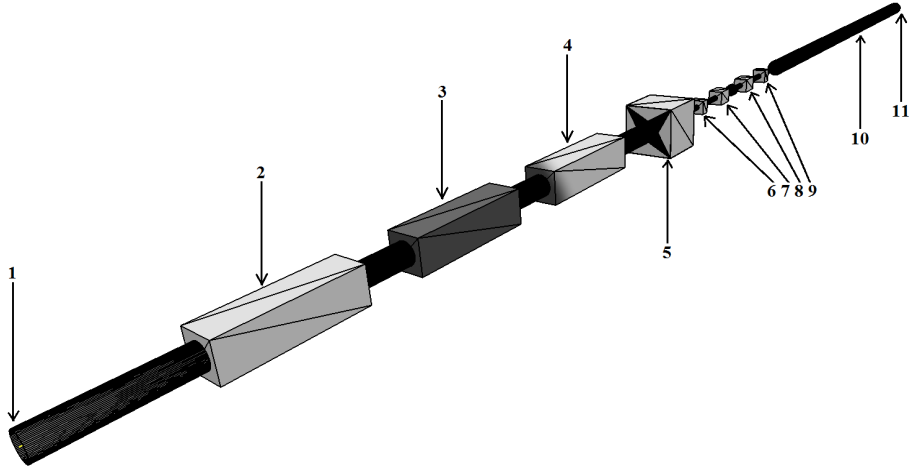}
\end{center}
\caption{3D view of the model of proton microscope made in Geant4:
1---beam source, 2--4---MQ1--MQ3 quadrupole lenses, 5---target chamber, 6--9---MQ4--MQ7 quadrupole lenses, 10---beam line pipe, 11---detector.}
\end{figure}

The ion optics of the facility includes the quadrupole magnetic lenses, which are grouped in two sections with various functions. The matching section consisted of three lenses (MQ1--MQ3). It provided optimum parameters of a proton beam before the investigated object: the size and angular distribution of a beam in the target plane. Quadrupole lenses MQ4--MQ7 were combined in the imaging--magneficating section to formation of the image of the object in the image plane, the magnification coefficient was accepted equal 4.1. The target with areal density up to 5~g/cm$^2$ was placed between matching and imaging sections. The Fourier's plane was forming in the center of imaging section between MQ5 and MQ6 quadrupole lenses. Protons cross this plane on various distances from the beam axis depending on the angle of the multiple Coulomb scattering gathered in the target. The larger angle of the proton scattering in the target  resulted in the greater distance of passing the proton from the beam axis  at Fourier's plane. A contrast of proton radiographic images was adjusted by installion, at Fourier's plane, a collimators with different aperture (angular aceptance). Protons with the angle of scattering more than angular aceptance of collimator were absorbed in the collimator body and did not participate in formation of the image. Joint adjustment of both sections of magnetic lenses is allowing to minimize of chromatic aberrations of proton radiographic images that considerably improved the quality of the images. The parameters of the ion-optical scheme of proton microscope have been numerical simulated by COSY Infinity code \cite{cosy} and are presented in table 1. The sign in parentheses designates polarity of the magnetic field. The proton source was modeled as point-source. The energy of protons was 247 MeV with energy spread $\Delta E/E =10^{-3}$, the number of protons in beam pulses was set $5 \times 10^8$, the beam spot had elliptic angular distribution (13.642~mrad in horizontal direction and 7.864~mrad in vertical direction). The following parameters of the magnetic quadrupole lenses were used: MQ1--MQ3 with an aperture 150~mm and 900~mm long, MQ4, MQ7 with an aperture of 40~mm and 80~mm long, MQ5, MQ6 with an aperture of~40 mm and 160~mm long. The collimator was modelled by the tungsten cylinder 40~mm length with a radius of 5~cm with an internal elliptic hole. The result of beam dynamic numerical simulation is presented in figure~2.

\begin{figure}[t]
\begin{center}
\includegraphics[width=0.8\columnwidth]{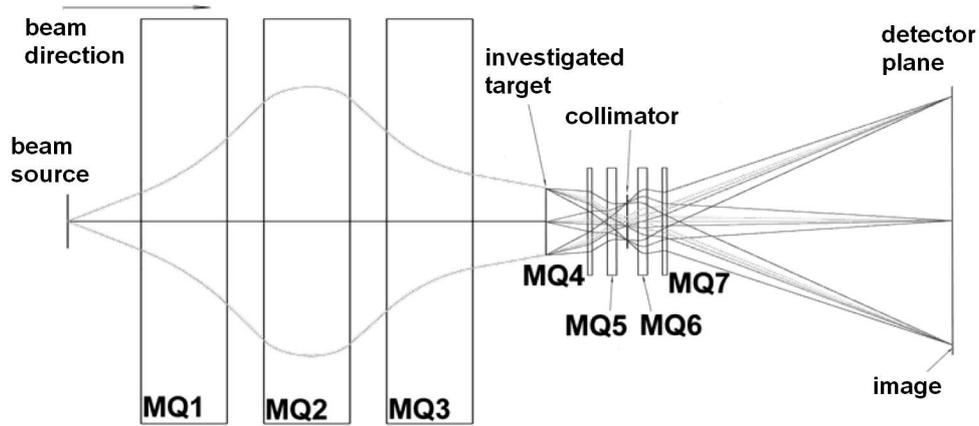}
\end{center}
\caption{The beam dynamic calculated in COSY Infinity code for 247--MeV proton microscope.}
\end{figure}

\Table{\label{Table_1}Parameters of the proton microscope.}
\br

Dimensions of installation, m\\

\mr
Distance to studied object&5\\
Distance after studied object&7\\
\mr
Magnetic field value at pole of quadrupole lenses, T\\
\mr
MQ1($-$), MQ3($-$)&0.215\\
MQ2(+)&0.266\\
MQ4(+), MQ5($-$), MQ6(+), MQ7($-$)&0.600\\
\br
\endTable

\Table{\label{Table_2}Parameters of the collimator.}
\br

Target material& PMMA& copper& gas\\
Angular acceptance, mrad & 1.5&8.0& 4.8 \\
Major semiaxes of the hole, mm & 2.73 & 13.04& 7.00\\
Minor semiaxes of the hole, mm & 1.70 & 7.66 & 5.50\\

\br
\endTable

\section{Results of Monte-Carlo numerical simulation}

The full-scale of the Monte-Carlo simulation of experiments was performed with developed, at Geant4 toolkit, virtual model of proton microscope. Optimum parameters of ion optics of proton microscope were calculated by COSY Infinity code when proton beam with energy 247~MeV passed through a thin object. Parameters of the facility were selected to provide the maximum sharpness of the image with better density sensitivity (contrast sensitivity). The collimator parameters providing optimum contrast sensitivity for different targets modeled in the work are specified in the table 2. The first type of investigated static targets were the step wedges made of a set of copper and polymethyl methacrylate (PMMA) plates with thickness from 250 to 2000 $\mu$m with a step of 250 $\mu$m. The results of Monte-Carlo simulation are shown in figure~3 and figure~4. Thickness of step wedges increases from the left to the right, i.e. than thicker is the object, the darker is its image.

\begin{figure}[t]
\begin{center}
\begin{minipage}{16pc}
\includegraphics[width=14pc]{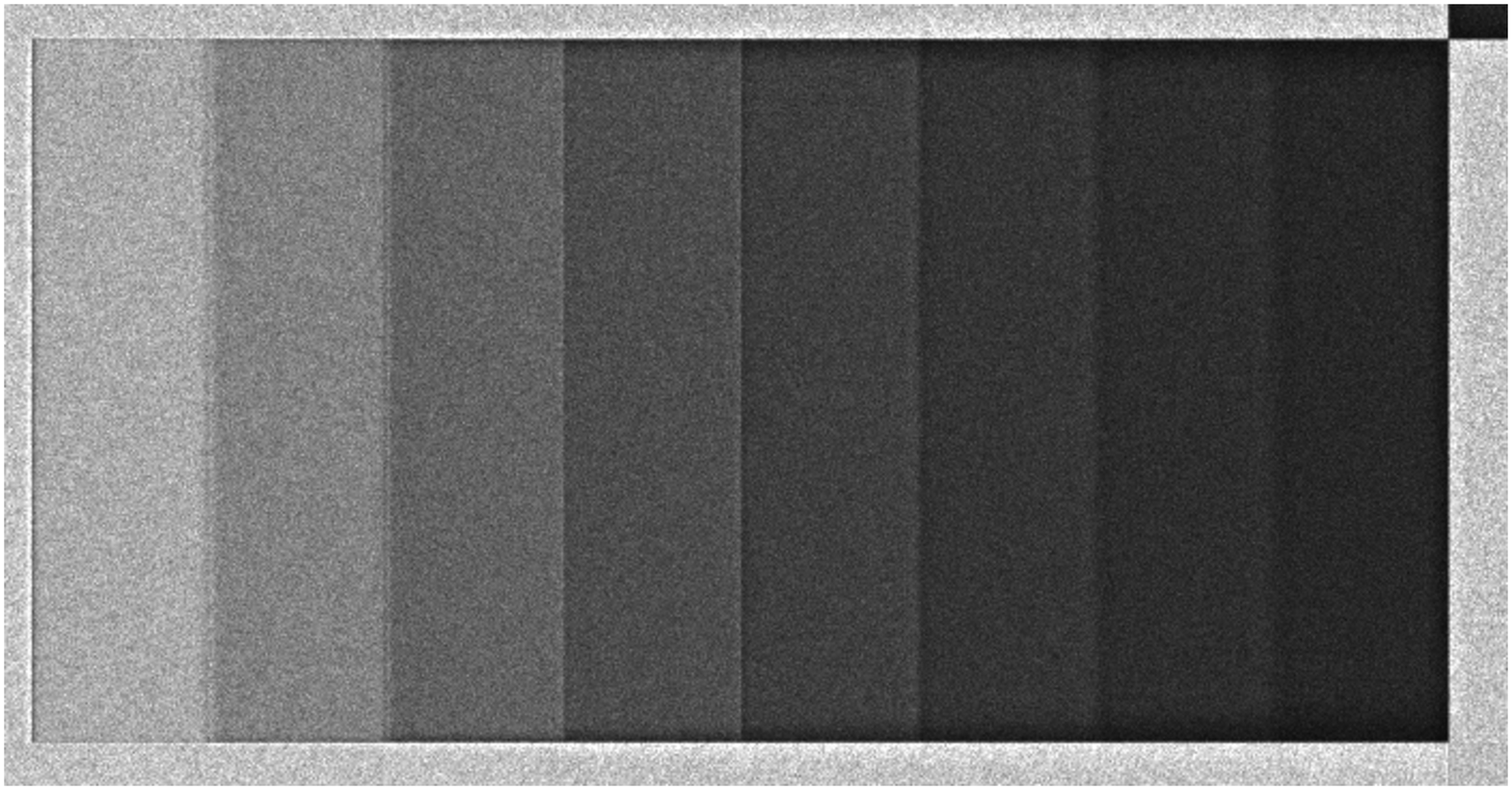}

\end{minipage}
\hspace{2pc}
\begin{minipage}{18pc}
\includegraphics[width=18pc]{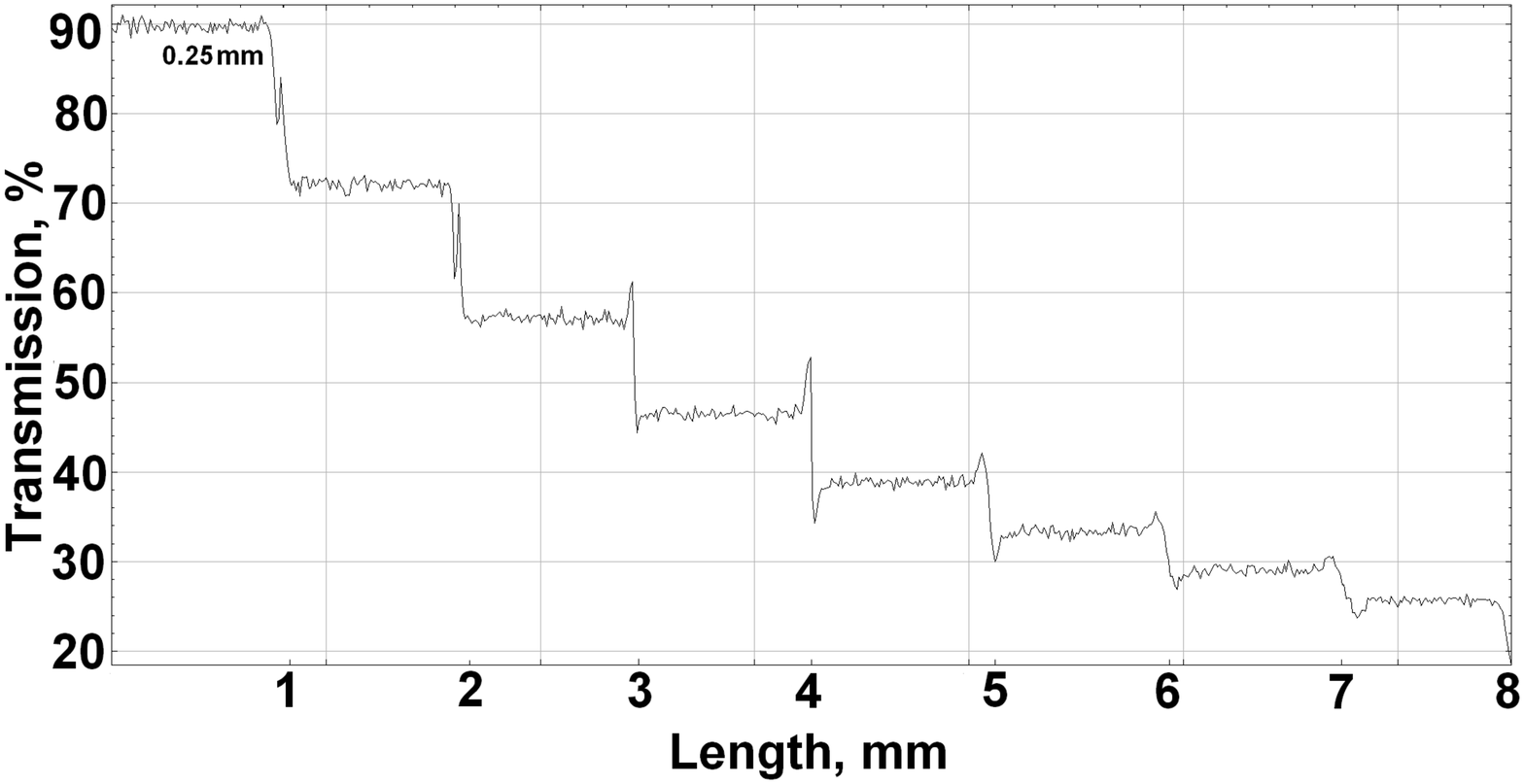}

\end{minipage}
\end{center}
\caption{Radiographic image of a copper step wedge (on the left) and a cross profile of its transmission, expressed as a percentage (on the right).}
\end{figure}

\begin{figure}[t]
\begin{center}
\begin{minipage}{16pc}
\includegraphics[width=14pc]{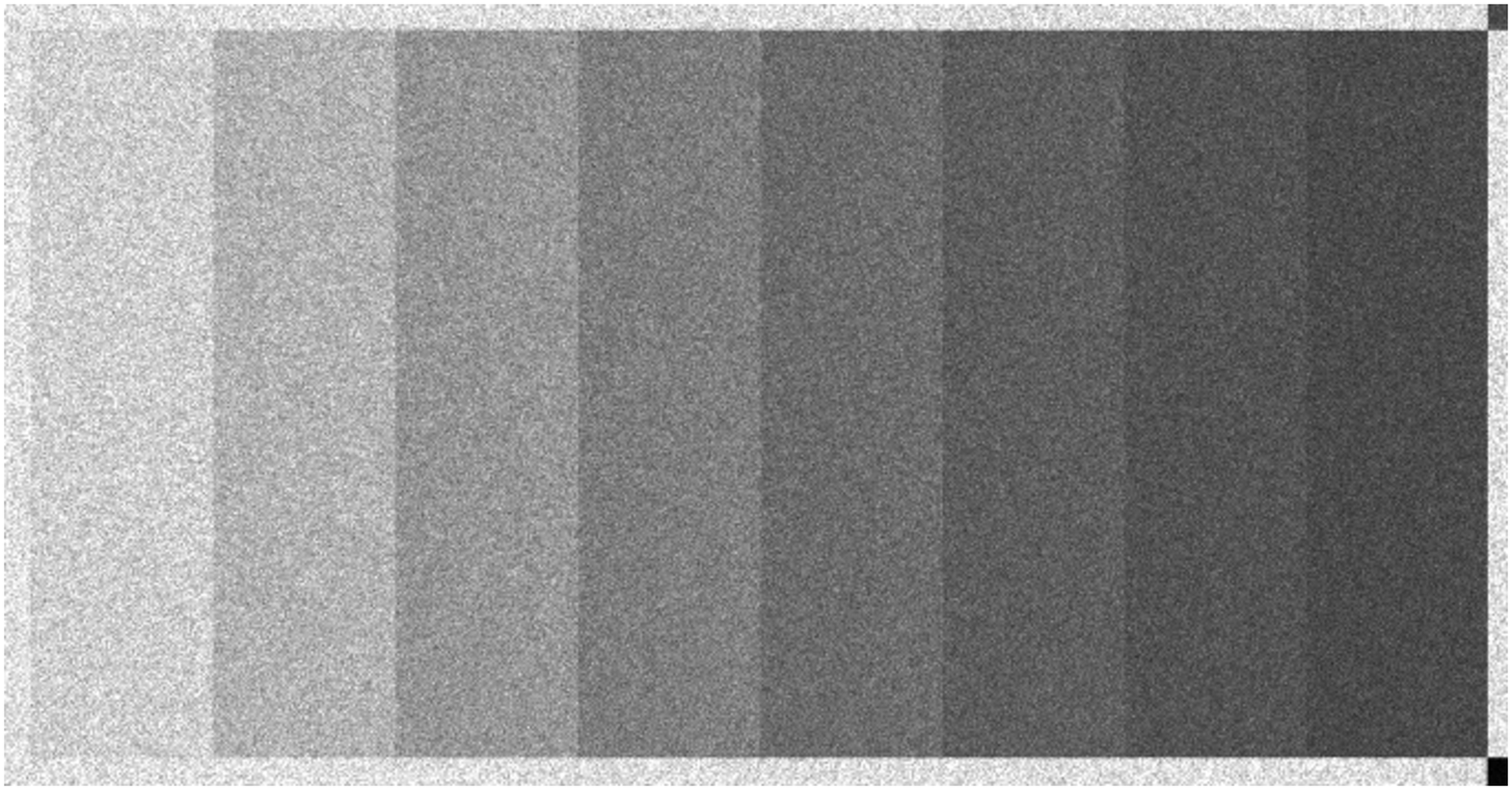}

\end{minipage}
\hspace{2pc}
\begin{minipage}{16pc}
\includegraphics[width=18pc]{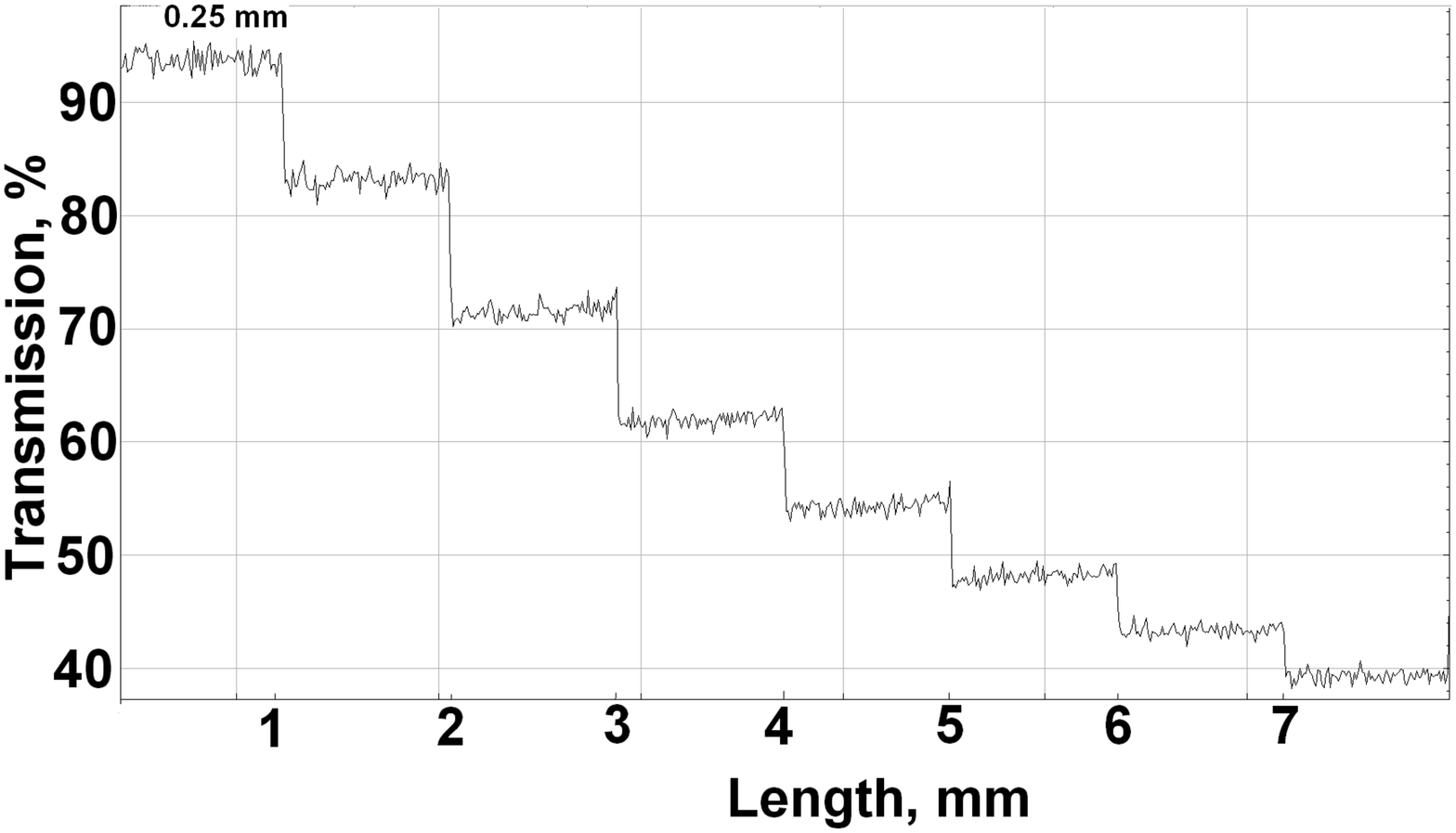}

\end{minipage}
\end{center}
\caption{Radiographic image of a PMMA step wedge (on the left) and a cross profile of its transmission, expressed as a percentage (on the right).}
\end{figure}

Radiography of shock-compressed xenon plasma by protons with energy 800~MeV has been performed earlier at PUMA proton microscope, some preliminary results were published in \cite{Kolesn2012}. The geometry of explosive target is shown in figure~5. The results of modeling of a static slug of xenon plasma with properties corresponding to plasma in experiment \cite{Kolesn2012} are presented in this work. The target was modeled by a cylindrical polypropylene tube with an internal diameter 22 mm and an external diameter 25 mm filled with gaseous xenon at pressure of 2.5~bars. The cylindrical area of the compressed gas 5 mm thick was located in the central part of the tube. The value of coefficient of compression was set by 7. Exact parameters of the shock-compressed plasma could be estimated by the measured mass velocity of xenon plasma and known Hugoniot adiabat of Xe \cite{Gryaznov80}. The thin wire markers were placed in the model of target for designation of the target axis. The simplified setup of an experiment was used in modeling: the target was described as a static object, effects of a curvature of the shock front were not taken into account, products of detonation and a copper foil were replaced with gaseous xenon with the initial pressure. It has been founded that optimum contrast sensitivity was realized at the collimator with the angular aceptance of 4.8~mrad. The region of the image with 100\% transmission corresponds to zero areal density. The transmission of uncompressed gas on the target axis is equal to 49\%, the transmission of compressed gas on the target axis is 16\%. In this case the change of transmission (not compressed gas-the compressed gas) is about 33\% that corresponds to the accuracy of determination of density better than 1\% if the registering device has dynamic range 1000. The model image of the shock-compressed gas is shown in figure~6.

\begin{figure}[t]
\begin{center}
\begin{minipage}{16pc}
\includegraphics[width=16pc]{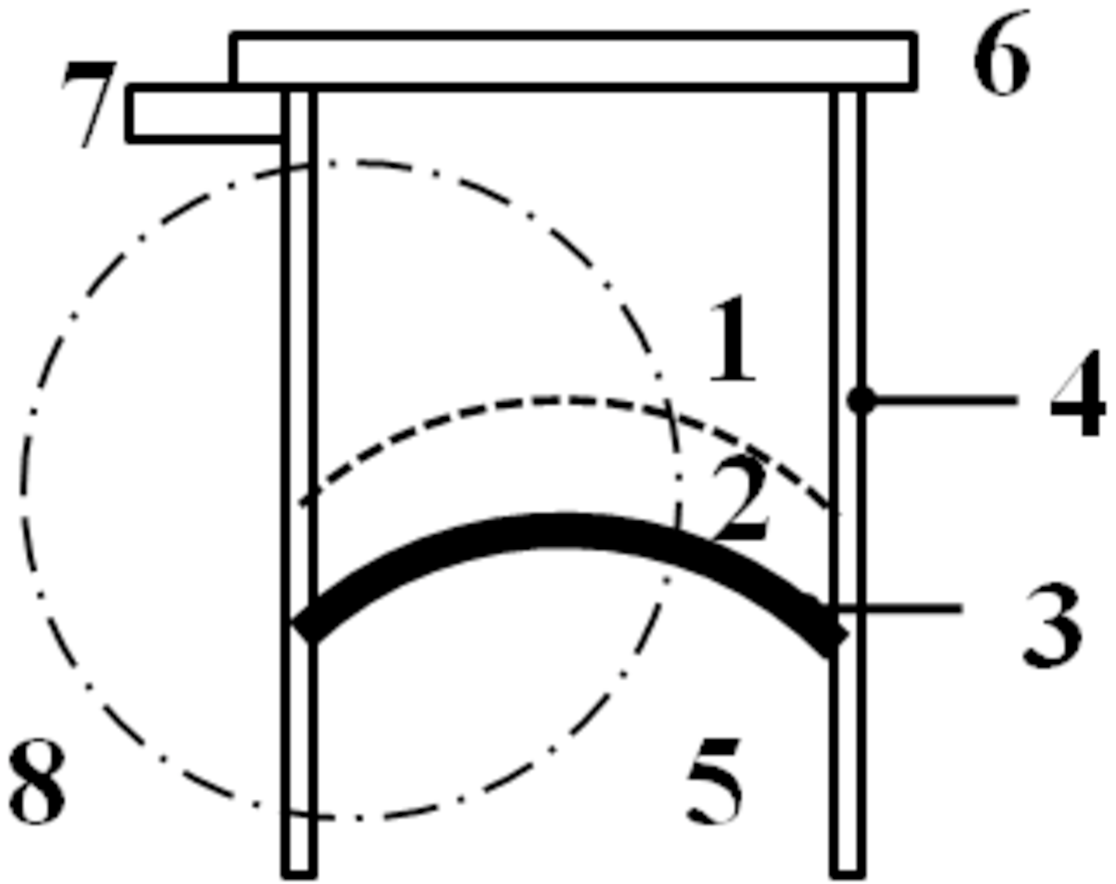}
\caption{Target scheme:
1---uncompressed gas, 2---shock-compressed xenon, 3---copper foil, 4---shell of generator, 5---products of detonation, 6---flange, 7---gas inlet.}
\end{minipage}
\hspace{2pc}
\begin{minipage}{16pc}
\includegraphics[width=16pc]{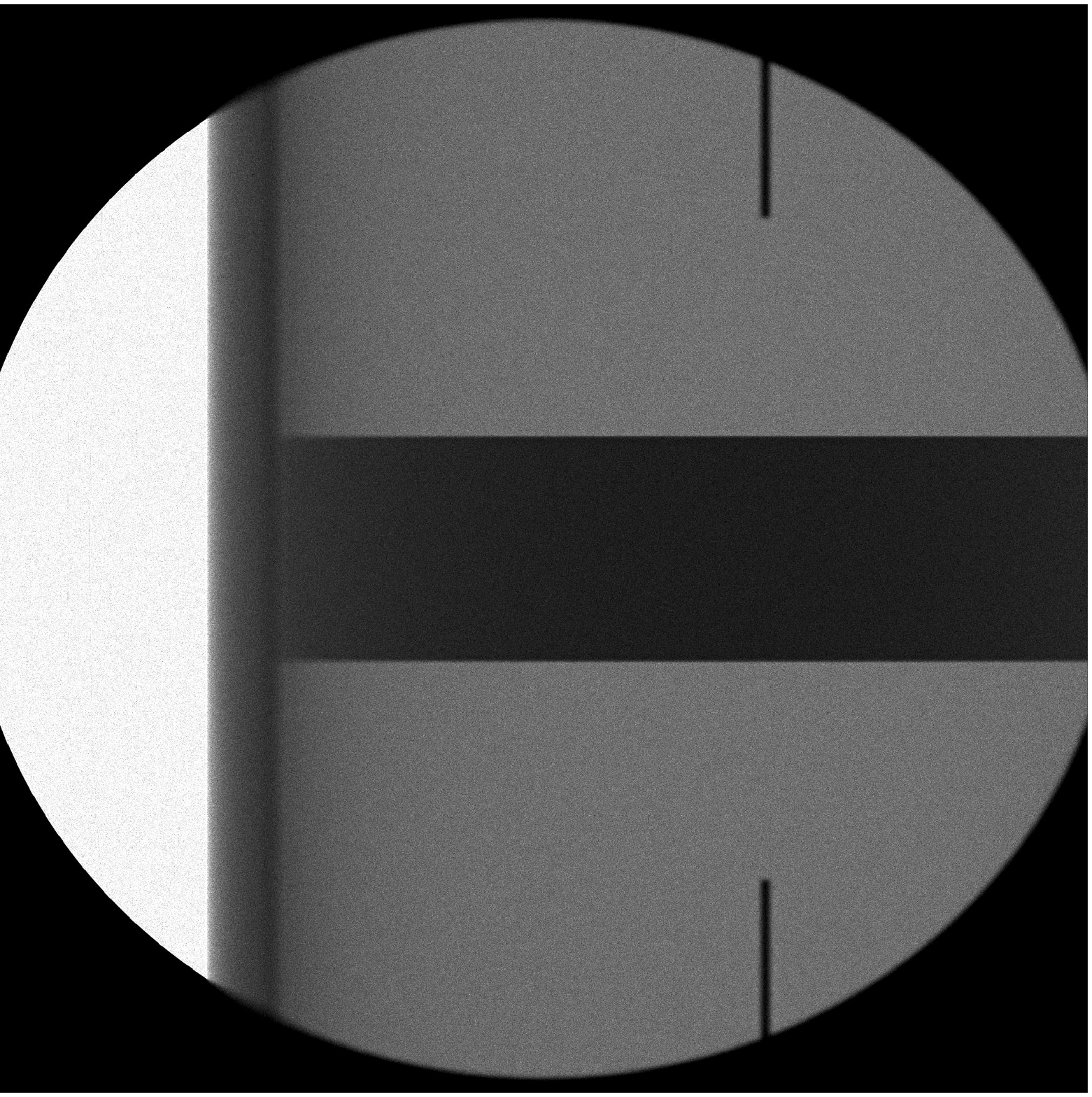}
\caption{Calculated radiographic image of simplified target.}
\end{minipage}
\end{center}
\end{figure}

\section{Conclusions}

The full-scale Monte-Carlo numerical simulation of radiographic experiments with developed virtual model of 247--MeV proton microscope were performed. The images of copper and PMMA static targets with variable thickness 250--2000 $\mu$m and image of shock-compressed xenon plasma with density about 0.1 g/cm$^3$ were simulated. The results of numerical simulation showed that in a dynamic experiment on proton microcope the density could be measured with accuracy better than 1\% in the objects with areal density about 5 g/cm$^2$.

\ack
This work was performed under financial support of the Russian Foundation for Basic Research (grant 15-08-09030), the Fair--Russia Research Center and the Presidium RAS (program 11P).

\section*{References}
\bibliographystyle{iopart-num}
\bibliography{kantsyrev_01}

\end{document}